\documentstyle[12pt]{article}

\def\bs{\begin{subequations}}
\def\es{\end{subequations}}

\catcode`\@=11

\newtoks\@stequation
\def\subequations{\refstepcounter{equation}
   \edef\@savedequation{\the\c@equation}%
   \@stequation=\expandafter{\theequation}
   \edef\@savedtheequation{\the\@stequation}
   \edef\oldtheequation{\theequation}%
   \setcounter{equation}{0}%
   \def\theequation{\oldtheequation\alph{equation}}}

\def\endsubequations{\setcounter{equation}{\@savedequation}%
   \@stequation=\expandafter{\@savedtheequation}%
   \edef\theequation{\the\@stequation}\global\@ignoretrue}

\catcode`\@=12

\catcode`\@=11
\def\vereq#1#2{\lower3pt\vbox{\baselineskip1.5pt \lineskip1.5pt
\ialign{$\m@th#1\hfill##\hfil$\crcr#2\crcr\sim\crcr}}}
\catcode`\@=12

\makeatletter
         \renewcommand{\theequation}{\thesection.\arabic{equation}}%
         \@addtoreset{equation}{section}%
\makeatother

\renewcommand{\thefootnote}{\fnsymbol{footnote}}

\begin{document}
\begin{titlepage}

August 1, 2002
\begin{center}        \hfill   \\
             \hfill     \\
                                 \hfill   \\

\vskip .25in

{\large \bf Experiment and Theory in Computations of
the He Atom Ground State \\}

\vskip 0.3in

Charles Schwartz\footnote{E-mail: schwartz@physics.berkeley.edu}

\vskip 0.15in

{\em Department of Physics,
      University of California\\
      Berkeley, California 94720}

\end{center}

\vskip .3in

\vfill

\begin{abstract}
Extensive variational computations are reported for the ground
state energy of the non-relativistic two-electron atom. Several 
different sets of
basis functions were systematically explored, starting with the original
scheme of
Hylleraas.  The most rapid convergence is found with a combination of
negative powers and a logarithm of the coordinate $s = r_{1}+ r_{2}$.
At N=3091 terms we pass the previous best calculation
(Korobov's 25 decimal accuracy with N=5200 terms) and we stop at N=10257 with
E = -2.90372 43770 34119 59831 11592 45194 40444 $\ldots$

Previous mathematical analysis sought to link the
convergence rate of such calculations to specific analytic properties
of the functions involved. The application of that theory to this new
experimental data leaves a rather frustrating situation, where we seem
able to do little more than invoke vague concepts, such as
``flexibility.''  We
conclude that theoretical understanding here lags well behind the
power of available computing machinery.

\end{abstract}

\vfill

\end{titlepage}

\renewcommand{\thefootnote}{\arabic{footnote}}
\setcounter{footnote}{0}
\renewcommand{\thepage}{\arabic{page}}
\setcounter{page}{1}

\section{Introduction}

For thousands of years mathematicians have enjoyed competing with
one another to compute ever more digits of the number $\pi$.
Among modern physicists, a close analogy is computation of the
ground state energy of the Helium atom, begun 75 years ago by E. A. Hylleraas
\cite{Hyll}.

Many have contributed incremental steps in this endeavor,
flexing their computational muscle and ingenuity, often trying
to use mathematical insight for advantage.  The strongest line of
theory has been to focus on the analytic properties of the
wavefunction, especially following the 1954 work of V. A. Fock
\cite{Fock} which showed a weak
logarithmic singularity at the three-particle coalescence.

The recent work of V. I. Korobov \cite{Koro} stands out for its
simplicity and its success.  His trial functions use
the three scalar coordinates
packaged as $e^{-\alpha r_{1}-\beta r_{2}-\gamma r_{12}}$,
with many sets of the complex nonlinear parameters $\alpha,
\beta, \gamma$  selected in a quasirandom manner within specified
intervals $A_{i} \le \alpha_{i} \le B_{i}$, etc. With a linear
combination of N=2200 terms of this type, grouped into four
  sets $[A_{i},B_{i}]$, Korobov surpassed the best previous work by
three decimals of accuracy in the Helium ground state energy; and
more recently he went on to N=5200 and added four more decimal places.

What struck me as surprising in Korobov's work was the fact that it
seemed to ignore that earlier ``wisdom'' about analytic
properties of wavefunctions.  His basis functions are, analytically,
no different from the original Hylleraas basis; yet his computational
results appear to converge so much more rapidly.  This perception
motivated the investigations reported below.

Section 2  presents the new experimental data - systematic
variational calculations using a variety of different basis
functions that have been suggested over the years: these include negative
powers, fractional powers and logarithms of the coordinates.
Vastly different rates of convergence are observed, as illustrated in
Figure 1.

Section 3  offers a qualitative discussion and attempts to interpret
this wealth of new data.  Section 4 is a review, and an attempt to apply,
the theoretical approach for understanding, at least semi-quantitatively,
the observed convergence rates.  We conclude that this state of the
theory is far from satisfactory.

\section{Experiments - Data}

Several different sets of basis functions were used in the standard
variational calculations for the ground state energy of the
Hamiltonian (with Z=2),
\begin{equation}
H = -\frac{1}{2} [\bigtriangledown_{1}^{2} + \bigtriangledown_{2}^{2}]
  -Z/r_{1} -Z/r_{2} + 1/r_{12} \label{b}
\end{equation}
and they are detailed below in terms of the Hylleraas coordinates,
\begin{equation}
s = r_{1} + r_{2}, \;\;\;\;\; t=r_{1} - r_{2}, \;\;\;\;\; u=r_{12} =
|\vec{x}_{1}-\vec{x}_{2}|.
\end{equation}

Table I presents summary results for the primary bases studied.  The
Accuracy of any energy value E is defined as Log[E*/(E*-E)] where
E* is our best estimate of the exact value.  Thus, the numerical
value of Accuracy tells how many correct decimal places there are in the
calculated result. Generally, we see that convergence was found to be
more rapid as one progressed through this series, A,B,C,D,E,F.  More
detailed discussion of the results will be deferred to the following
sections.

$\bullet$ Basis A is just the original Hylleraas set:
\bs
\begin{eqnarray}
\psi = \sum C_{l,m,n}\; e^{-ks/2}\; s^{l}\;u^{m}\;t^{n} \label{a} \\
l,m = 0,1,2,3,\ldots , \;\;\; n=0,2,4,6,\ldots
\end{eqnarray}
\es
and we use the order parameter $\omega = l+m+n$ to group the
terms.  We also designate a calculation of order $\omega$ to mean a
basis set including all terms with $l+m+n \leq \omega$. The total
number of terms, N, grows asymptotically as $\omega^{3}/12$. This same
organizational scheme is used for all the experiments listed below.

$\bullet$ Basis B allows negative powers of s, which were introduced by
Kinoshita \cite{Kino}:
\begin{equation}
e^{-ks/2}\; s^{l}\;(u/s)^{m}\;(t/s)^{n}; \label{d}
\end{equation}
and we can rearrange the terms as
\begin{equation}
e^{-ks/2}\; s^{l}\;(u/s)^{m+n}\;P_{n}(t/u),
\end{equation}
using the Legendre polynomials to get the selection rule $\Delta n =
0, \pm 2$. This technique follows the work of
Goldman \cite{Gold} and leads to more efficient use of computer space
and time.

$\bullet$ Basis C allows fractional powers of s, which were first introduced by
H.~M.~Schwartz \cite{Sch0}:
\begin{equation}
(1, s^{1/2})\;e^{-ks/2}\; s^{l}\;u^{m}\;t^{n}
\end{equation}
which doubles the size N of the basis for each order $\omega$. (The
$s^{1/2}$ term is omitted for $\omega=0$.)

$\bullet$ Basis D uses the logarithm of s, first introduced by
Frankowski and Pekeris \cite{Fran}:
\begin{equation}
(1, ln(s))\; e^{-ks/2}\; s^{l}\;u^{m}\;t^{n}
\end{equation}
(The ln(s) term is omitted for $\omega = 0,1$.  The values of N shown
in Tables I and II are two units off for bases D and F.)

$\bullet$ Basis E has both negative powers and fractional powers of s.
\begin{equation}
(1, s^{1/2})\;e^{-ks/2}\; s^{l}\;(u/s)^{m}\;(t/s)^{n}.
\end{equation}

$\bullet$ Basis F has both the logarithm and negative powers of s.
\begin{equation}
(1, ln(s))\;e^{-ks/2}\; s^{l}\;(u/s)^{m}\;(t/s)^{n}.
\end{equation}

For each of the basis sets described above, the scale parameter k was
varied somewhat to find the lowest energy at each order.  For the
Table I data on Basis A, k varied from 5.0 to 8.2; for Basis B, from 
3.8 to 5.9;
for Basis C, from 4.7 to 6.6; and for Basis D, from 4.6 to 6.6.  For
Bases E and F, it was found that the optimum value of k stayed close
to 2.0 for the mid-size and larger orders, so k was fixed at this value for
all the data shown.

In two cases I replaced the set of functions $e^{-ks/2}\;s^{l}$ by the
set $e^{-\alpha_{l}s}$, using Korobov's quasi-random method for
selecting the (real) nonlinear parameters $\alpha$ in a single group.
The results for these
experiments are shown as Bases B' and E' in Table I.

Several variants of these basis sets were also explored briefly but
discarded when they appeared less effective, as functions of N, than
their counterparts above.  Among these were the use of:
\bs
\begin{eqnarray}
&&\textup{Negative powers of s and u } \;(u/s)^{m}\;(t/u)^{n}   \label{c} \\
&&\textup{Third roots of s instead of the square root} \\
&&\textup{Two or more powers of } ln(s) \\
&&\textup{The coordinate }  r = \sqrt{r_{1}^{2} + r_{2}^{2}} \\
&&\textup{The coordinate }  R = |\vec{x}_{1} + \vec{x}_{2}| \;\;\;\;
\textup{ (See the Appendix.)}
\end{eqnarray}
\es
\newpage

\begin{tabular}{|r|r||c||c||c|}
\multicolumn{5}{l}{\textbf{Table I.} Energy Accuracy -- Summary data}\\
\hline
$\omega$ & N & A-Basis & B-Basis & B' -Basis \\ \hline
9 & 125 & 7.9 & 8.7 & 9.4 \\
12 & 252 & 8.7 & 10.2 & 10.7 \\
16 & 525 & 10.4 & 11.7 & 12.3 \\
21 & 1078 & 11.6 & 13.2  & 13.9 \\
27 & 2135 & 12.8 & & \\
\hline \hline
$\omega$ & N & C-Basis & D-Basis &  \\ \hline
7 & 139 & 9.5 & 9.6 & \\
9 & 249 & 11.2 & 11.3 & \\
12 & 503 & 13.4 & 13.5 & \\
16 & 1049 & 15.8 & 16.0 & \\
21 & 2155 & 18.3 & 18.5 & \\
\hline \hline
$\omega$ & N & E-Basis & E' -Basis & F-Basis \\ \hline
7 & 139 & 9.5 & 9.8 & 9.4 \\
9 & 249 & 11.7 & 11.6 & 11.5 \\
12 & 503 & 14.5 & 14.5 & 14.3 \\
16 & 1049 & 18.2 & 18.3 & 18.1 \\
21 & 2155 & 22.5 & & 22.4 \\
27 & 4269 & 27.2 & & 27.6 \\
34 & 8093 & 29.7 & & 33.0 \\ \hline
\end{tabular}

Accuracy = Number of correct decimals
\vskip 1cm

In Table II are the detailed results (in a compact format)
for the two bases - E and F -
that show the most rapid convergence. One quantitative measure of the rate of
convergence is the Ratio of successive differences and this is also
shown in the table.

\vskip 1cm

Technical notes.  For these computations, I wrote a set of subroutines
for multiple-precision arithmetic (in C), eschewing more professional
packages which are available.  The last row of data in Table II used
  101 decimals of precision and took one week running on an
otherwise idle
desktop computer equipped with a 300 MHz processor and 320 MB of memory.

\newpage

\begin{tabular}{|r|r||l|c||l|c|}
\multicolumn{6}{l}{\textbf{Table II.} Calculated results with the two 
best bases} \\
  \hline
  $\omega$ & N & E-Basis Energies & Ratios & F-Basis Energies & Ratios \\ \hline
  4 & 43 & -2.90372 183 &  & -2.90371 941 & \\
  5 & 67 & -2.90372 42300 & 17.9 &-2.90372 415 & 22.5 \\
  6 & 99 &  '' 43643 & 11.3 & -2.90372 43610 & 14.2 \\
  7 & 139 & 43762 2 & 15.8 & '' 43758 7 & 14.0 \\
  8 & 189 & 43769 759 & 14.3 & 43769 382 & 12.2 \\
  9 & 249 & 43770 28348 & 10.5 & 43770 25283 & 11.2 \\
  10 & 321 & 43770 33352 & 7.32 & 43770 33068 & 8.35 \\
  11 & 405 & '' 34036 & 9.37 & '' 34000 4 & 8.80 \\
  12 & 503 & 34109 6 & 8.30 & 34106 294 & 8.94 \\
  13 & 615 & 34118 444 & 8.63 & 34118 13633 & 9.02 \\
  14 & 743 & 34119 46685 & 8.79 &34119 44847 & 10.1 \\
  15 & 887 & 34119 58229 & 8.06 &34119 57846 & 7.33 \\
  16 & 1049 & '' 59667 & 10.4 & '' 59620 & 10.0 \\
  17 & 1229 & 59806 & 6.32 & 59797 & 5.90 \\
  18 & 1429 & 59828 5 & 10.5 & 59827 3 & 9.38 \\
  19 & 1649 & 59830 654 & 5.12 & 59830 456 & 5.50 \\
  20 & 1891 & 59831 06419 & 9.69 & 59831 03831 & 8.89  \\
  21 & 2155 & 59831 10650 & 5.08 & 59831 10381 & 6.18 \\
  22 & 2443 & '' 11482 & 9.14 & '' 11442 & 8.22 \\
  23 & 2755 & 11573 8 & 5.62 & 11571 6 & 7.25 \\
  24 & 3093 & 11589 997 & 7.94 & 11589 408 & 6.66 \\
  25 & 3457 & 11592 03902 & 6.18 & 11592 08081 & 8.87 \\
  26 & 3849 & 11592 36947 & 5.29 & 11592 38154 & 4.79 \\
  27 & 4269 & '' 43186 & 4.80 & '' 44444 & 11.0 \\
  28 & 4719 & 44484 & 2.97 & 45017 4 & 3.57 \\
  29 & 5199 & 44922 & 2.88 & 45177 752 &13.1  \\
  30 & 5711 & 45074 35 & 2.32 & 45189 95689 & 3.00  \\
  31 & 6255 & 45139 97 & 2.29 & 45194 02040 & 14.8 \\
  32 & 6833 & 45168 66 & 2.16 & 45194 29566 & 2.76 \\
  33 & 7445 & 45181 95 & 2.11 & '' 39521 & 15.1 \\
  34 & 8093 & 45188 24 & 2.06 & 40182 & 2.78 \\
  35 & 8777 & 45191 29 & 2.02 & 40420 7 & 13.5 \\
  36 & 9499 & 45192 80 & & 40438 342 & 3.11\\
  37 & 10259 & & & 40444 00495 & \\ \hline
\end{tabular}

extrapolate: E* = -2.90372 43770 34119 59831 11592 45194 40444 6
\vskip 1cm
\nopagebreak

\section{Experiments - Discussion}

Figure 1 provides a visual comparison  of the convergence rates
for the different variational basis sets (A - F), plotting Accuracy
  versus the Log of N, the number of basis functions
used.  I found a number of surprises in these results.

Surprise 1.  Basis B (negative powers of s) shows a significant improvement
over basis A (the original Hylleraas basis).

Surprise 2.  Bases E and F do a great deal better than any of the
others.  Somehow, the benefits of B and C (or B and D) are cumulative.

Surprise 3.  The performance of C and D are nearly identical, as with
E and F (until we reach very high orders.)

Surprise 4.  The performance of basis E drops off dramatically after
  $\omega = 27$; but basis F keeps up its rapid convergence, although
with marked oscillation, as seen from the Ratios in Table II.

The surprising performance of Korobov's basis has already been noted: his
published results are shown by the line labelled with
  the letter ``K'' in Figure 1.

The fact that Basis C performs a lot better than Basis A was not a
surprise, since earlier work \cite{Sch1} had already shown that.  Ditto for
Basis D \cite{Fran}. Also, the smallness of the gain shown by basis B' over B
(and by E' over E) is as expected, based upon
the analytic equivalence of exponentials and power series.

While I cannot explain the surprising results, I can readily offer
suggestions on how one might interpret them. The virtue of Bases C and D
lies in providing more flexibility to the ``radial'' behavior of the
wavefunction (coordinate s); while that of the negative
powers lies in providing more flexibility in the ``angular''
behavior (ratios u/s and t/s). The similarity between C and D (and between E
and F) indicates that
the precise analytic behavior in the ``radial'' coordinate is not
important - any flexibility will do - until one gets to the
very high orders.

This appeal to ``flexibility'' is just armwaving; it lacks any
mathematical foundation.  Such appeal to flexibility is also the best way
I know to understand the success of Korobov's calculations: his work seems
akin to the ``adaptive'' techniques used in numerical integration,
where one puts additional mesh points into any region that shows a
slower rate of convergence.

In varying the scale parameter k, I most always found a simple minimum
in the energy. However, in one case - Basis E at $\omega = 21$ - a
more complex behavior emerged:  see Table III.  While the variation
shown here is not very great, this does raise the general question of how
effectively one may search for the minimum of a complicated function of
many nonlinear parameters.  This is a possible source of worry in
using Korobov's technique, especially when it comes to error
estimation and extrapolation.

\vskip 1cm

\begin{tabular}{|c||c|c|c|c|c|c|c|c|}
\multicolumn{9}{l}{\textbf{Table III.} Double minimum in E(k) for Basis E at
  $\omega$ = 21} \\ \hline
k & 1.7 & 1.8 & 1.9 & 2.0 & 2.1 & 2.2 & 2.3 & 2.4 \\ \hline
E & 10567 & 10697 & 10691 & 10650 & 10635 & 10640 & 10636 & 10603 \\ \hline
\multicolumn{9}{l}{These numbers for the energy E follow the first 20 decimal
places.} \\
\end{tabular}

\section{Theories}

The first lesson in analysis of atomic wavefunctions concerns the
two-particle cusps: linear behavior as any one of the coordinates
$r_{1}, r_{2}$, or $r_{12}$ goes to zero.  All of the basis functions
studied in this paper are correct in that regard; we are concerned here
with what comes next.

Take the Hylleraas expansion (\ref{a}) and put it into the
Schrodinger equation (\ref{b}). Then collect the coefficients of each
monomial in s, u, and t and set that equal to zero.  Early in this
infinite set of algebraic equations for the expansion coefficients
$C_{l,m,n}$ one finds the following \emph{inconsistent} equations
\bs
\begin{eqnarray}
C_{1,0,0} + Z C_{0,0,0} = 0 \\
-2 C_{1,1,0} + C_{1,0,0} = 0 \\
4 C_{1,1,0} - C_{1,0,0} = 0.
\end{eqnarray}
\es

This contradiction in the Hylleraas basis was discovered by Bartlett,
Gibbons and Dunn \cite{Bart}
  in 1935 and it led them to consider an alternative expansion: one
that involved logarithms of the hyperradius $r = \sqrt{r_{1}^{2} +
r_{2}^{2}}$.  Later, Fock \cite{Fock} independently developed a systematic
expansion of the wavefunction with such terms.

In 1962 this author \cite{Sch2} developed a general theory about the 
convergence
rate of variational calculations, based upon analogy with
least-squares fitting of functions and one-dimensional model problems.
This line of analysis was expanded by others \cite{Morg} and in some
cases given a  more rigorous mathematical basis. \cite{Hill}

That theoretical work led directly to the idea that the convergence rate in
  Hylleraas-type
calculations for the Helium ground state was controlled by the Fock logarithmic
singularity; and the semi-quantitative analysis seemed to fit the available
data.  It also led to the successful exploitation of the fractional
power basis C \cite{Sch1}.  Shortly thereafter, Frankowski and
Pekeris \cite{Fran} took
logarithmic terms explicitly into their trial functions and this also
seemed to confirm the importance of the Fock behavior.

However, when Kinoshita \cite{Kino} considered the systematic use of negative
powers - in the form (\ref{c}), not (\ref{d}) - he found that there was no
contradiction of the type noted above.  I have confirmed that this
conclusion holds also for the basis (\ref{d}) used in the current work.

Thus, one might be drawn to believe that the logarithmic singularity is not
an intrinsic property of the He atom wavefunction itself but rather
results from a particular analysis that relies on the six-dimensional
coordinate r.  This idea may be dispelled by reading the work of
Myers et al \cite{Myer}, where they take one for a walk around the
three-particle coalescence and show that the (finite) discontinuity in the
local energy disappears when one includes the full set of terms that accompany
the Fock logarithm.

This approach helps us understand some other experimental results.
We noted earlier that inclusion of negative powers $t/u$ gave poorer
results than $t/s$.  Each of these ratios shows a (finite) discontinuity
when one walks around the place where the denominator
vanishes -  something that the correct wavefunction should
not allow.  In the  case of (t/u) this error occurs along a line,
while in the  case of (t/s) this error occurs only at a point.
A similar situation appears in the work of Goldman \cite{Gold}, who holds the
previous record for basis size (N=8066). His use of
the coordinates $r_{<}, r_{>}$ led to a very efficient computational
scheme, but demonstrates relatively slow
convergence. (See the point ``G'' in Figure 1.)  This may be
attributed to the discontinuity of his basis functions on the 2-dimensional
surface $r_{1}= r_{2}$.

\subsection{Fitting the Data}
The goal of a good theoretical understanding should be the ability to
predict or to explain, at least semi-quantitatively, the observed
rates of convergence for systematic variational calculations with different
basis sets.  In my earlier work \cite{Sch2}, the attempt to do this 
was based upon
analogies with one-dimensional model problems, doing least-squares fit
with appropriate orthogonal bases to represent functions
with various types of singularities:
\bs
\begin{eqnarray}
Minimize && \int \rho(x) dx\;[f(x) - \sum_{i=0}^{n-1} C_{i} u_{i}(x)]^{2} \\
&& C_{i} = \int \rho(x) dx f(x) u_{i}(x) \\
&& Error \approx (C_{n})^{2}.
\end{eqnarray}
\es
For one example we find:
\begin{eqnarray}
f(x) = x^{\nu}\; ln x ,&&\;\;\;  \rho(x) = x^{\mu}\;\;\;\;\;
  \textup{on the interval (0,1)} \nonumber \\
C_{n}& \sim& 1/n^{\mu + 2\nu + 3/2}; \label{p}
\end{eqnarray}
and an alternative example is:
\begin{eqnarray}
f(x) = x^{\nu}\;ln x, &&\;\;\;  \rho(x) = x^{\mu}e^{-x}\;\;\;\;\;
  \textup{on the interval (0,$\infty$)} \nonumber \\
C_{n}&\sim& 1/n^{\mu/2 + \nu + 1}. \label{q}
\end{eqnarray}
The difference in convergence rates for these two examples may be
understood qualitatively as follows.  The basis functions
$x^{n}e^{-x}$ peak at $x=n$. Therefore, at higher n these basis
functions on the interval $(0,\infty)$ get farther and farther away
from the singularity, which is at $x=0$.  One may improve the
situation by using basis functions $x^{n}e^{-kx}$, where k is a scale
parameter that may grow as one proceeds to higher orders.  I do not
have a quantitative theory for this result but it is qualitatively
relevant to the current study.

In my 1962 work I applied this simple modeling to the He atom problem,
identifying the Fock term $r^{2} ln r $ as the dominant singularity
which is neglected in conventional Hylleraas coordinates. This led me
to predict a convergence rate formula,
\begin{equation}
E(\omega) - E(\omega -1) \sim const./\omega ^{p} \label{m}
\end{equation}
and I estimated that p should be between 5.5 and 10, due to
uncertainties in replacing the real 3-dimensional problem with the
one-dimensional model.  The then best results with Hylleraas
variables (work of Pekeris, 1959, up to order 21, using a cleverly
orthogonalized basis) fit the convergence rate formula (\ref{m}) with 
a value of
p between 7 and 8.  This was good confirmation of the theory.  The
extended computations reported here (Basis A data in Table I) fit the
convergence rate formula (\ref{m}) with a value of p which varies from 7, at
the lower orders, to a value about 12 at the higher orders.  This
improvement is probably due to my allowing the scale parameter k to
vary, which was not done in the earlier work.

Also, in 1962, I introduced the half-powers of coordinate s,
explicitly for the purpose of increasing the convergence rate,
following this theory.  That was successful, with the observed value of p
approximately doubled to 14 or 15 at $\omega \le 8$. The extended
computations reported here (Basis C data in Table I) are fit to
values of p which vary from about 16 to 21.  Again, this is fairly
good confirmation of the theory; and again we acknowledge some
improvement by allowing the scale parameter to vary.

Following that earlier theory one would certainly not expect Basis D
to converge at the same rate as Basis C -- but this is exactly the
behavior we have found in the present experiments.

What can I say about the observed convergence rate of Basis B,
introducing negative powers into the Hylleraas functions?  The data
in Table I are fit with a value of the exponent p around 13.  I do not
understand this but will only offer a guess that it may have to do
with fitting the complex ``angular'' behavior around the Fock
singularity, which was described in Ref.~\cite{Myer}. Maybe this is connected
with the difference in
convergence rates noted above, in Eqs. (\ref{p}, \ref{q}), for the model
problems on (0,1) and on
(0,$\infty$ ).

Finally, look at the results for Bases E and F.  The
data in Tables I and II are fit with values of the exponent p which grow from
the 20's to the 40's in the middle range of $\omega$;  at the top end,
the data for Basis E drop to around p=25, while the data for Basis F climb to
about p = 65.  I am at a loss to explain these large exponents
following the former analysis.

An alternative to the power law convergence rate formula (\ref{m}) is the
exponential rate formula
\begin{equation}
E(\omega) - E(\omega -1) \sim const. (\sigma)^{\omega} \label{n}
\end{equation}
which one could expect from a model fitting problem that involved no
singularities at all. For example, expanding $e^{-ax}$ in terms of
$x^{n}\; e^{-bx}$  would yield the formula (\ref{n}) with $\sigma =
(\frac{a-b}{a+b})^{2}$.  If one plots the data for Basis F (Log of
increments in E vs.
$\omega$), it does look close to a straight line; and the smoothed
data in Table II may be fitted
with a value of $\sigma$ in the range 0.13 - 0.16 for  $\omega > 16$.  If
one looks at the asymptotic behavior of the He wavefunction as $r_{1}$
goes to infinity, the behavior in $r_{2}$ should be as $e^{-Zr_{2}}$
with $Z=2$.  The trial functions I used for this basis have the
exponential envelope $e^{-k(r_{1}+r_{2})/2}$ with $k=2.$ Using the
formula quoted above, this model gives us  the
parameter $\sigma$ as $(\frac{2-1}{2+1})^{2} = 0.11$.  This looks
like a fairly good fit to the data;
but accepting this explanation
would lead us to doubt the relevance of the Fock singularities for the He
wavefunction.

Also, I know of no published theoretical attempts to explain  the excellent
convergence found by Korobov with his highly nonlinear fitting of
the trial wavefunction.  John Morgan has suggested (in private communication)
  that Korobov's approach may be likened to the work of
fitting the Hydrogen radial wavefunction with a set of gaussians,
using ``floating exponents''\cite{Kutz}.  This sounds plausible, but
at present it is just more handwaving about ``flexibility.''.

I conclude that theoretical understanding of the convergence of
variational calculations on the two-electron atom is far outstripped
by the raw computing power of available machinery.

Some may ask if any of this is really relevant to current issues in
physics. One response is to point to high accuracy measurements
performed on atomic systems which may check the current theories of fundamental
particles and interactions.  A recent paper by
Pachucki and Sapirstein \cite{Pach}  aims to determine the fine structure
constant to a few parts-per-billion.  This is based upon  measurements of
the $2^{3}P_{J}$ states in Helium and
detailed calculations that rely upon a Korobov-type
representation of the atomic wavefunction.

Then, again, all this may be nothing more than an expression of
$\pi$-envy.

\vspace{1cm}

\noindent {\bf ACKNOWLEDGEMENT}

I am grateful to John D. Morgan III for several very helpful discussions.

\vskip 1.0cm
\setcounter{equation}{0}
\def\theequation{A.\arabic{equation}}
\boldmath
\noindent{\bf Appendix: Integrals}
\unboldmath
\vskip 0.5cm

Integrals of the following type were needed in the calculations
reported here:
\begin{equation}
\int_{0}^{\infty} ds \; e^{-s}\; s^{p}\; (ln(s))^{q}.
\end{equation}
There is a simple recursion formula on the index p; and for the
minimum values of p I used a particular technique of numerical
integration. (See Ref. \cite{Sch3}.)  First change variables, $ s = 
exp(y)$; then use
the simple rule,
\begin{equation}
\int_{-\infty}^{\infty} f(y)\; dy \approx h\;
\sum_{n=-\infty}^{\infty} \; f(nh).
\end{equation}
The summation is truncated when terms are smaller than the desired
accuracy; and the answer converges exponentially as the interval $h$
is decreased.

\vskip 0.5cm

For the two-electron atom, one can evaluate the most conventional
integrals from the formula,
\begin{equation}
  \int \frac{d^{3}x_{1}}{4\pi}\;\int \frac{d^{3}x_{2}}{4\pi}\;\;
\frac{e^{-ar_{1}}}{r_{1}}\;\frac{e^{-br_{2}}}{r_{2}}\;\frac{e^{-cr_{12}}}
{r_{12}}\;
  = \frac{1}{(a+b)(b+c)(c+a)}, \label{e}
  \end{equation}
and derivatives of this simple result with respect to the parameters a,b,c.
\vskip 1cm
In exploring more complicated functions, I was able to find another simple
formula for
the following integral, which involves $R = |\vec{x}_{1} + \vec{x}_{2}|$.
\begin{eqnarray}
\int \frac{d^{3}x_{1}}{4\pi}\;\int \frac{d^{3}x_{2}}{4\pi}&&\;
\frac{e^{-ar_{1}}}{r_{1}}\;\frac{e^{-br_{2}}}{r_{2}}\;\frac{e^{-cr_{12}}}
{r_{12}}\; \frac{e^{-dR}}{R} \label{f} \\
=&&\frac{1}{(a^{2}+b^{2}-2c^{2}-2d^{2})} \;
ln\frac{(a+b+2c)(a+b+2d)}{2(a+c+d)(b+c+d)}. \nonumber
  \end{eqnarray}
To derive this, insert the Laplacian operators into the middle of the
integral and let them work both ways.  It appears that one could
almost deduce these results (\ref{e}) and (\ref{f}) purely by arguments of
analyticity and symmetry. Consider, for example, how the integral behaves
  as $r_{1} \rightarrow \infty$: by counting powers one sees the nature
  of the singularity as (a+c), or (a+c+d), goes to zero.

  As noted earlier, using this variable R in the He trial
  wavefunction did not produce good results - as one might expect since
  it introduces a spurious cusp when the two electrons are on opposite
  sides of the nucleus.  I have, nevertheless, recorded the above
  information here in case it might be useful to others.

  The result (\ref{f}) can be generalized with $R = |\mu\vec{x}_{1} +
  \nu\vec{x}_{2}|$.

\newpage

\newpage

\begin{picture}(360,420)(0,0)
\thicklines
\put(0,0){\framebox(350,400){}}
\thinlines
\put(0,50){\line(350,0){350}}
\put(0,100){\line(350,0){350}}
\put(0,150){\line(350,0){350}}
\put(0,200){\line(350,0){350}}
\put(0,250){\line(350,0){350}}
\put(0,300){\line(350,0){350}}
\put(0,350){\line(350,0){350}}
\put(-14,-4){ 0}
\put(-14,46){ 5}
\put(-14,96){10}
\put(-14,146){15}
\put(-14,196){20}
\put(-14,246){25}
\put(-14,296){30}
\put(-14,346){35}
\put(-14,396){40}
\put(-8,-14){125}
\put(50,0){\line(0,4){4}}
\put(42,-14){250}
\put(100,0){\line(0,4){4}}
\put(92,-14){500}
\put(150,0){\line(0,4){4}}
\put(140,-14){1000}
\put(200,0){\line(0,4){4}}
\put(190,-14){2000}
\put(250,0){\line(0,4){4}}
\put(240,-14){4000}
\put(300,0){\line(0,4){4}}
\put(290,-14){8000}
\put(330,-14){N (Log scale)}
\put(40,430){Accuracy (the number of correct decimal digits)}
\put(60,415){vs. N (the number of basis functions)}
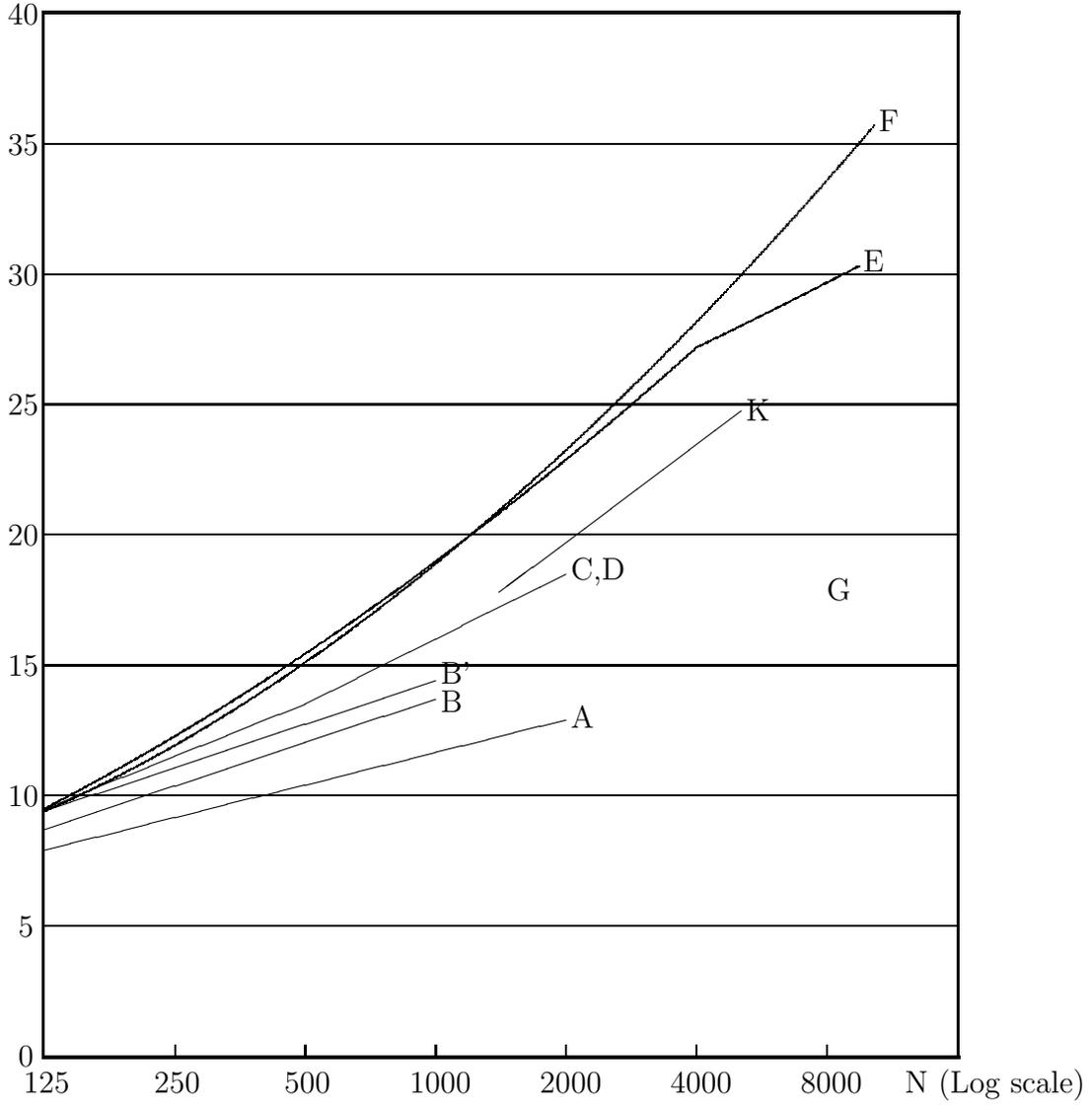
\put(50,-70){\textbf{Figure 1.}  Comparative Convergence Rates}
\put(0,79){\line(4,1){200}}
\put(202,126){A}
\put(0,87){\line(3,1){150}}
\put(152,132){B}
\put(0,94){\line(3,1){150}}
\put(152,143){B'}
\put(200,185){\line(-2,-1){100}}
\put(100,135){\line(-5,-2){100}}
\put(202,183){C,D}
\qbezier[500](0,95)(125,160)(250,272)
\qbezier[100](250,272)(281,287)(312,303)
\put(314,301){E}
\qbezier[600](0,94)(150,160)(318,357)
\put(320,355){F}
\put(267.3,247.6){\line(-4,-3){93.1}}
\put(269,244){K}
\put(300,175){G}

\end{picture}

\end{document}